\documentclass[MNA]{mna}

\usepackage{xcolor}
\usepackage{float}
\restylefloat{figure}
\usepackage{enumerate}
\usepackage{comment}

\SHORTTITLE{Sub-Riemannian model of neural states in M1}

\TITLE{A sub-Riemannian model of neural states
  in the primary motor cortex}

\AUTHORS{%
  Caterina~Mazzetti\footnote{Dipartimento di Matematica, Universit\`a di Bologna.
    \EMAIL{caterina.mazzetti2@unibo.it}}
  \and
  Jawad~Ali\footnote{Dipartimento di Matematica, Universit\`a di Bologna.
    \EMAIL{jawad.ali13@studio.unibo.it}}
  \and
  Alessandro~Sarti\footnote{Centre d’Analyse et de Math\'ematique Sociales, Sorbonne Universit\'e.
    \EMAIL{alessandro.sarti@ehess.fr}}
  \and
  Giovanna~Citti\footnote{Dipartimento di Matematica, Universit\`a di Bologna.
    \EMAIL{giovanna.citti@unibo.it}}}

\SHORTAUTHOR{Mazzetti, Ali, Sarti, Citti}


\AMSSUBJ{53C17 ; 92B20 ; 62H30}

\SUBMITTED{January 10, 2025}
\ACCEPTED{February 18, 2026}

\ARXIVID{2501.03247}

\VOLUME{6}
\YEAR{2026}
\PAPERNUM{1}
\DOI{10.46298/mna.15057}

\ABSTRACT{We develop a neurogeometric model for the arm area of
motor cortex, which encodes complex motor primitives, ranging from
simple movement features like movement direction, to short hand
trajectories, termed fragments, and ultimately to more complex
patterns known as neural states (Georgopoulos, Hatsopoulos,
Kadmon-Harpaz et al). Based on the sub-Riemannian framework
introduced in 2023, we model the space of fragments as a set of
short curves defined by kinematic parameters. We then introduce a
geometric kernel that serves as a model for cortical connectivity
and use it in a differential equation to describe cortical activity.
By applying a grouping algorithm to this cortical activity model, we
successfully recover the neural states observed in Kadmon-Harpaz et
al, which were based on measured cortical activity. This confirms
that the choice of kinematic variables and the distance metric used
here are sufficient to explain the phenomena of neural state
formation. The modularity of our model reflects the brain's
hierarchical structure, where initial groupings in the kinematic
space $\mathcal{M}$ lead to more abstract representations. This
approach mimics how the brain processes stimuli at different scales,
extracting both local and global properties.}

\begin{document}


\section{Introduction}

Our study aims to model the functional architecture of motor cortical cells used to
control arm reaching movements.

Pioneering works in motor cortical research were developed by A. Georgopoulos, 
who first recorded the cell's selectivity properties  (see \cite{georgopoulos1984static, kettner1988primate} and \cite{georgopoulos1982relations, schwartz1988primate}). According to his studies cell's response is maximal when the hand position and direction coincide with a position and direction, characteristic of the cell.  More recently it has been proved that the tuning for movement parameters is not static, but varies with time (\cite{ashe1994movement, moran1999motor, churchland2007temporal, paninski2004spatiotemporal}); Hatsopoulos in \cite{Encoding, reimer2009problem} proposed that individual motor cortical cells rather encode ``movement fragments", i.e. short trajectories of the hand.
Each fragment is characterized as a 
trajectory with approximately constant direction, 
and with speed increasing up to a maximum or decreasing to a minimum. In \cite{kadmon2019movement}, applying a grouping algorithm to the cortical activity measured of this family of cells, the cells (hence the fragments) were clustered in eight classes, but the authors could not recover the same grouping, using a distance defined in terms of the kinematic properties of the fragments.  

Mathematical models for the description of hand trajectories and based on optimality principles have been proposed in various papers (\cite{todorov2006optimal, hogan1984organizing, FH, uno1989formation,  flash2007affine,  biess2007computational, flash2001computational, jean2010optimal}). 
The approach followed in these articles is the setting of a nonholonomic control system, whose underlying structure is defined in terms of sub-Riemannian geometry (see F. Jean's book \cite{jean2014control}). 

A different approach was proposed in \cite{mazzetti2023functional}, inspired by neurogeometric models of the visual cortex (see \cite{hoffman1970higher,petitot1999vers,citti2006cortical}) and by movement perception in visual areas \cite{cocci2015cortical}). The authors assume that a motor neuron can be represented by a point 
\begin{equation}\label{M}(x, y,t, \theta, v, a) \in M:=\mathbb{R}_{(x,y)}^2  \times \mathbb{R}_{t} \times S_\theta^{1} \times \mathbb{R}^+_{v} \times \mathbb{R}_{a},\end{equation} where $(x, y)$ denotes the hand's position in a two-dimensional plane, $t$ denotes the time. The velocity vector will be represented in polar coordinates by two variables: $\theta$  which denotes the direction of velocity and $v$ which denotes the absolute value of the speed in this direction. Finally, $a$ denotes the acceleration in the direction $\theta$ so that it will be represent as a scalar. Using the differential constraints operating on the variables (see \eqref{6D} below) the authors were able to introduce a sub-Riemannian structure and a distance in the space $M$ defined in \cite{mazzetti2023functional}, and refer to \cite{agrachev2019comprehensive, ledonne2021lecture} for a general presentation).  
Fragments were formally recovered as short admissible curves in this structure. In addition to applying a grouping algorithm in this space, the authors were able to decompose a trajectory in fragments \cite{mazzetti2023functional}. However, the fragments were not yet clustered in neural states. 

\medskip

Our scope is to model the organization in neural states experimentally found in \cite{kadmon2019movement}.  Indeed they do not only observe that neurons in  $\mathcal{M}$ are sensitive to hand trajectory, but they also cluster the elementary trajectories in so-called neural states. A first model trajectories clustering was presented in \cite{mazzetti2023direction}: to each elementary trajectory it is associated its mean orientation and
acceleration, and the grouping is performed in these variables. Though efficient, the algorithm does not seem to be neurally implemented, since there is no experimental evidence that the neurons compute means over the fragments. On the contrary, it seems that neurons code properties of fragments evolving in time,  hence we work in this space of curves with values of $\mathcal{M}$ which models the space $\mathcal{F}$ of fragments. 
We also remark that the classification of  \cite{kadmon2019movement} is invariant with respect to the spatial variables, hence we introduce a sub-manifold $\mathcal{M}_1$ of the manifold $\mathcal{M}$, which is independent of the position $(x,y)$. The notion of sub-Riemannian sub-manifold has been introduced in \cite{franchi2001rectifiability} and  \cite{franchi2007regular} (see also \cite{Ambrosio2006intrinsic} and \cite{citti2006implicit} for the expression of the vector fields induced on the sub-manifold). We will use their approach and the estimate of \cite{nagel1985balls} to find the distance induced on $\mathcal{M}_1$ by the immersion in  $\mathcal{M}$. With this distance, we will introduce a pseudo-metric in the space $\mathcal{F}$ of curves with values $\mathcal{M}$, which is the space of fragments. A spectral clustering with this metric will allow to recover the clustering obtained in \cite{kadmon2019movement}. 

Let us explicitly recall that the clustering in \cite{kadmon2019movement} is based only on the neural activity, and the authors were unable to obtain the same classification with a distance based only on kinematic variables. On the contrary our classification is based only on a kinematic model. This proves that the choice of these variables is sufficient to explain this phenomenon and the distance we consider is consistent with cortical connectivity. 
In addition we are using a clustering algorithm in the space of fragments, obtained by a previous grouping. This modular approach seems to be the correct instrument to describe the functionality of the brain able to describe visual or motor input at different scales.

The structure of the paper is the following. In Section \ref{sec2} we present in detail the experiment of \cite{kadmon2019movement} which we want to model and we recall the neurogeometrical model of \cite{mazzetti2023functional}. In Section \ref{sec3}, we introduce our geometric model of neural states, expressed by a grouping algorithm. In Section \ref{sec4}, we apply our algorithm it to artificially generated and to real data, and we compare the neural states found with the present kinematic model with the one of \cite{kadmon2019movement} obtained from neural data. 
Section \ref{sec5} contains the conclusions.

\section{The state of the art}\label{sec2}

\subsection{Motor cortex functionality: features, fragments, neural states}
It has been experimentally proved that neurons in the motor cortex are sensitive to progressively more complex motor primitives: from simple features, as direction of movement, to short trajectories of the hand, called fragments, to more complex patterns, which are neurally implemented in the so called neural states.

\textbf{Features coded in motor areas} The first studies of the motor cortex were due to A. Georgopoulos, who recognized that motor neurons code the direction of movement trajectory (see \cite{georgopoulos1982relations}) . After that it has been proved that neurons are sensitive to other features which reflect kinematic properties of movement, such as position, time, velocity and acceleration of the hand both in two-dimensional and three-dimensional space (see \cite{kalaska2009intention, ashe1994movement, georgopoulos1984static}). 

\textbf{Fragments}
Later on, Hatsopoulos \cite{Encoding} (see also \cite{reimer2009problem}) highlighted that tuning to movement
parameters varies with time and proposed to describe the activity of neurons through
a trajectory encoding model, that is as a composition of fragments. In particular, fragments are characterized an accelerating or decelerating phase and almost constant direction of movement. 
Churchland and Shenoy \cite{churchland2007temporal}  proposed an analogous model which describes the temporal
properties of motor cortical responses. 

\begin{figure}[H]
\centering
	\includegraphics[scale= 0.35]{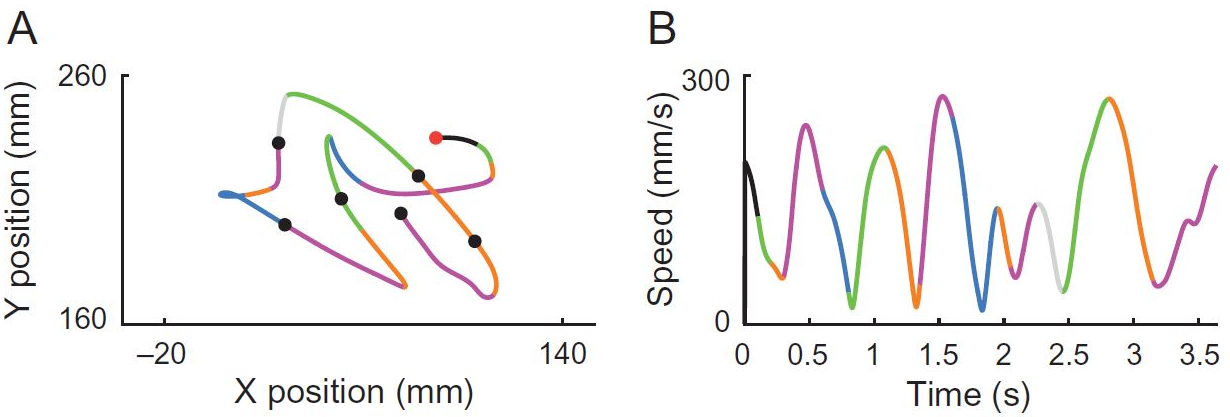}
\caption{Here, the random target pursuit task is represented: the starting point of the motion is the red dot, and subsequent targets are black circles.  The hand trajectory is represented by two images: (A) represents position in a 2D plane, and (B) represents the speed profile. Movement is segmented into fragments, which are characterized by almost constant orientation (see (A)) and acceleration or deceleration phase (B). Finally, each color represents a single neural state. Image taken from \cite{kadmon2019movement}.}
\label{fig2}
\end{figure}

Since neurons are sensitive to kinematic parameters, a trajectory of the hand is considered as a curve in the space of position, direction of movement, velocity and acceleration. 
It is often visualized through two images: the projection in the plane of the $(x,y)$ position variables, where one can also estimate the direction of movement and one in the plane of the time and speed variables $(t,v)$: the tangent to the graph allows to evaluate the acceleration (see Figure \ref{fig2}). 

\begin{figure}[H]
\centering
\includegraphics[scale= 0.43]{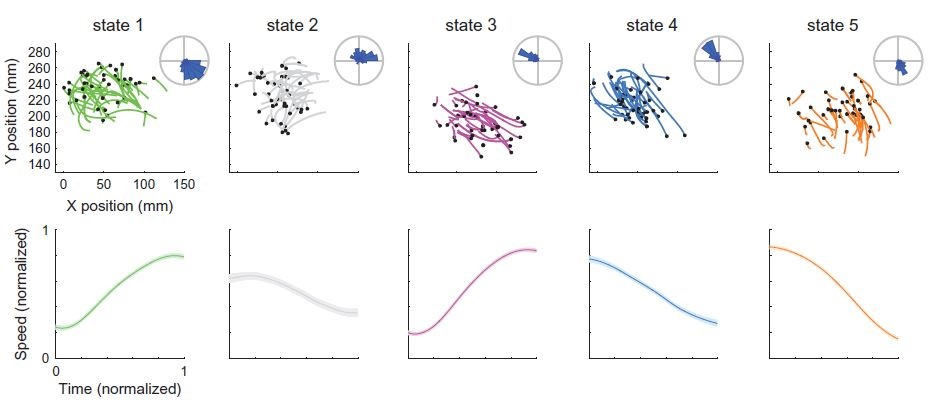}
\caption{Clustering of fragments in  neural states obtained in \cite{kadmon2019movement}. The two images visualized in each column represent the fragment: above is represented it $(x,y)-$ section and below a normalized profile of the $(t, v)-$section). Radial histograms show the mean directions of all the trajectories within each neural state. }
\label{clusterini}
\end{figure}

\textbf{Neural states} Starting from the paper  \cite{graziano2002cortical}, it became clear that neuron in the primary motor cortex M1 are sensitive to even more complex pattern. 
In 2019, N. Kadmon Harpaz, D. Ungarish, N. Hatsopoulos and T. Flash \cite{kadmon2019movement} studied the activity of neural populations in the primary motor cortex of macaque monkeys during a random-target pursuit (RTP) task and a center-out reaching task. The authors processed neural activity by identifying sequences of coherent behaviours, called neural states, by means of a Hidden Markov model\cite{kemere2008detecting}.  
In addition to decomposing movement, the obtained fragments were grouped and each group called neural state(see Figure \ref{clusterini}). Each of these identifies a group of fragments all with comparable direction in the $(x,y)$ plane and with a specific 
acceleration and deceleration phase in the $(t,v)$ plane. 
The obtained neural states did not show selectivity to movement speed and amplitude. 

The movement segmentation and the clustering into states was obtained from neural data, and one of the problem posed by the authors of \cite{kadmon2019movement} was to find a distance able to recover the same clustering only by using kinematic variables. 

\subsection{Mathematical models of movement fragments}

\subsubsection{A model of the feature space}

A first kinematic model of the decomposition of movement in fragments was proposed in \cite{mazzetti2023functional}.
The authors considered that motor cortical cells are sensitive to 
 hand's position in a two-dimensional plane,  time, direction of movement, velocity and acceleration. Consequently the motor cells were identified by 6 components $(x,y,t,\theta, v, a)$, where 
the triple $\left(t, x, y\right)\in\mathbb{R}^3$, accounts for a specific hand's position in time, the variable $\theta\in S^1$ encodes hand's movement direction, and the variable $v$  represents hand's speed in this direction: in this way we can assume that  $\theta\in [0, 2\pi]$,   and $v \geq 0,$ so that
$(\theta, v)$ completely individuates the velocity vector. Finally $a$ denotes acceleration in the same direction $\theta$. The space of features was denoted 
\begin{equation}\label{6D}
\mathcal{M}= \mathbb{R}^{3}_{\left(t,x,y\right)} \times S^1_{\theta} \times \mathbb{R}^+_{v} \times \mathbb{R}_{a} . 
\end{equation}
The quantities selected as features are not independent, but 
they are related by differential constraints, which were expressed 
by means of the vanishing of suitable  1-forms. In particular, the variables on the basis: $(x, y, t)$ are independent, while $\theta$  is the direction of movement, and it satisfies  
$\frac{dy}{dx} = \frac{\sin(\theta)}{\cos(\theta)}$. This condition will be  expressed by imposing that vector fields belong to the kernel of the one-form
\[
\omega_{1} = -\sin(\theta) d x + \cos(\theta) d y= 0.
\]
According to its definition $v$ denotes the speed in the direction $\theta$ so that it satisfies $ v = \cos(\theta) \frac{d x}{dt} + \sin(\theta) \frac{d y}{dt}$ , which leads to impose the 1-form
\[
\omega_{2} = \cos\theta d x + \sin\theta dy - vd t= 0,
\]
Finally, from the definition of acceleration, $a = \frac{dv}{dt}$ we get 
\[\omega_{3} = d v -a\; dt= 0.\]

A possible choice of vector fields orthogonal to these forms $\omega_i$ 
is \footnote{More formally the operation formally analogous to a the scalar product between a form and a vector field is called duality. Precisely, if $\omega = \sum_{i=1}^n a_i dx_i $ is a 1-form and $X = \sum_{i=1}^n b_i {\partial{x_i}} $ is a vector field, we can duality $<\omega, X> = \sum_{i=1}^n a_i b_i $, and we say that $X$ belongs to the kernel of $\omega$ if $<\omega, X>=0$. 
}

\begin{align}\label{campi_2D}
X_{1}= v\cos\theta\frac{\partial}{\partial{x}} + v\sin\theta\frac{\partial}{\partial{y}}+ a\frac{\partial}{\partial{v}}+ \frac{\partial}{\partial{t}},\quad
X_{2}= \frac{\partial}{\partial{\theta}},\quad
X_{3}= \frac{\partial}{\partial{a}}.
\end{align}

\subsubsection{A differential model of fragments}

The choice of these vector fields, together with a metric which makes them orthonormal, introduces the sub-Riemannian manifold. By definition, horizontal curves are integral curves of the vector fields $X_1, X_2$ and $X_3$, and can be expressed as

\begin{equation}\label{curvaa}
\gamma'\left(s\right)= \alpha_1\left(s\right)X_{1}\left(\gamma\left(s\right)\right)+ \alpha_2\left(s\right)X_{2}\left(\gamma\left(s\right)\right)+ \alpha_3\left(s\right)X_{3}\left(\gamma\left(s\right)\right),\\
\end{equation}
where the coefficients $\alpha_1$ are not necessarily constants. The variable $s$ denotes a parametrization of $\gamma,$ so that it is not intrinsic. 
If $\alpha_1=0,$ this would imply that 
$t'=0,$ which is not meaningful. 
 In particular, curves with $\alpha_1=0$  and the other parameters different from $0$, would  change acceleration or direction of movement without changing in time.
 This is geometrically possible, but physically impossible. Consequently, $\alpha_1$ is always different from $0,$ and we can assume $\alpha_1=1$. This implies that  $t=s,$
and selects a subset of the horizontal curves, which we will call admissible curves. Then the equation for admissible curves reduces to

\begin{equation}\label{curvaa}
\gamma'\left(t\right)= X_{1}\left(\gamma\left(t\right)\right)+ \alpha_2\left(t\right)X_{2}\left(\gamma\left(t\right)\right)+ \alpha_3\left(t\right)X_{3}\left(\gamma\left(t\right)\right),\\
\end{equation}
that is the vector tangent to trajectories in $\mathcal{M}$.

 From a purely physiological point of view, we can assume that signal propagates only along admissible curves, which model the geometry of axons.
 
 In addition, in  \cite{mazzetti2023functional} 
the curves expressed in \eqref{curvaa} were proposed 
both as a model of connectivity, and 
 as a model of fragments. 
 More precisely the authors proved
that the full fan experimentally found in 
\cite{churchland2007temporal, Encoding} can be obtained 
as a set of curves $\gamma(t)$  solutions of equation\eqref{curvaa}, defined on an interval $[0,T]$, and with polynomial coefficients. The coefficients $\alpha_1$ and  $\alpha_2$ can be choosen to be constant, while the choice of $\alpha_3$ which ensures that the acceleration vanishes at the initial and final point and has a bell shaped graph is that \begin{equation}\label{eqa}a' = \alpha_3(t) = j\Big(t- \frac{T}{2}\Big),\end{equation}
where $j$ is a real number. 
\subsubsection{Fragments obtained via grouping in the sub-Riemannian space of features}

Since the signal propagates along axons of cells, and propagation has been described through admissible curves, distance in $\mathcal{M}$ is related to connectivity weights. We can naturally introduce in the cortical space $\mathcal{M}$ the associated Carnot Carath\'eodory distance (see for example \cite{citti2024horizontal}). 

\begin{figure}[H]
\centering
\includegraphics[scale= 0.8]{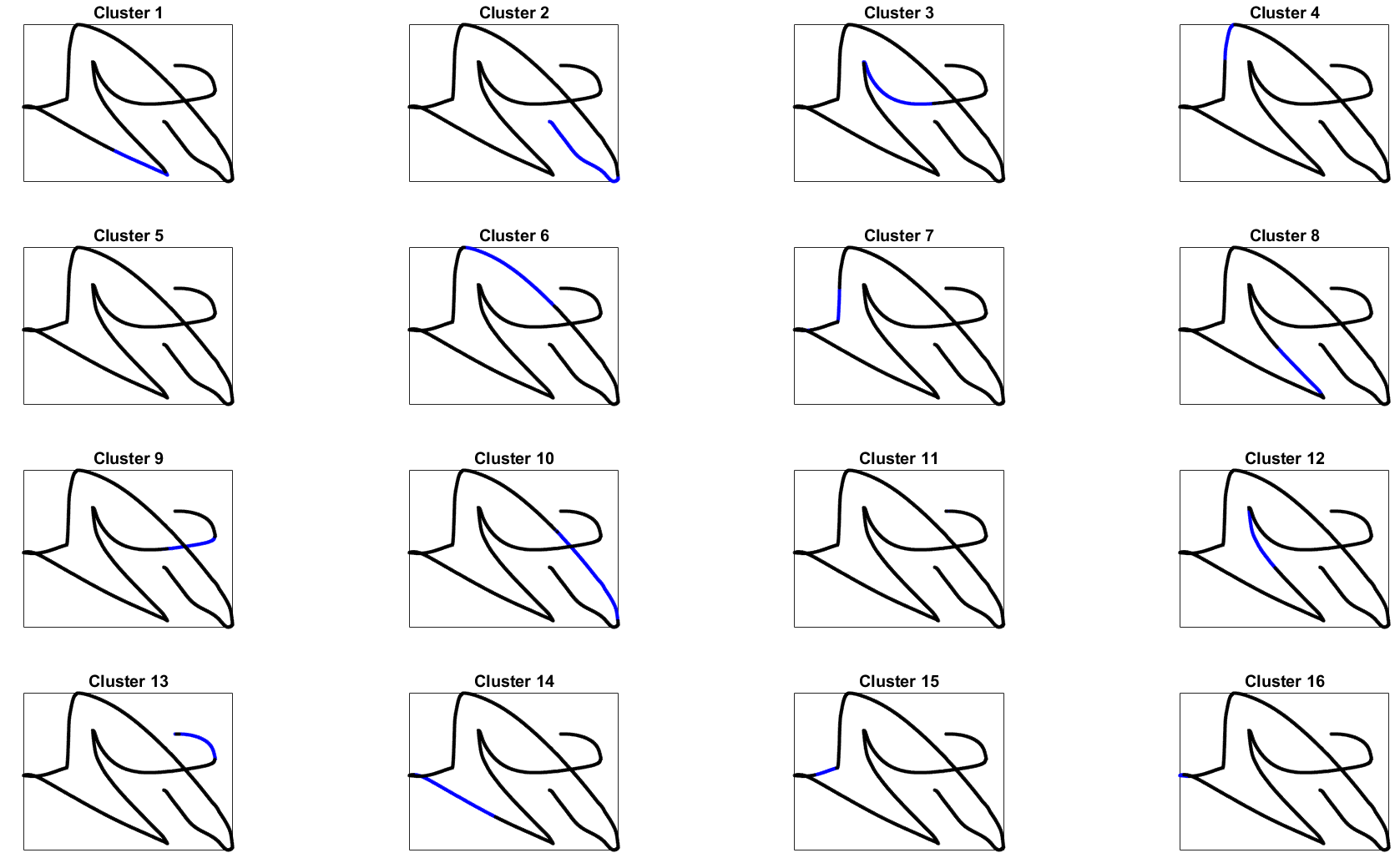}
\caption{Movement is decomposed in fragments of the $(x,y)$ components: fragments are not organized in  neural states. Source: \cite{mazzetti2023functional}.}
\label{clusterini_mazzetti1}
\end{figure}

\begin{figure}[H]
\centering
\includegraphics[scale= 0.25]{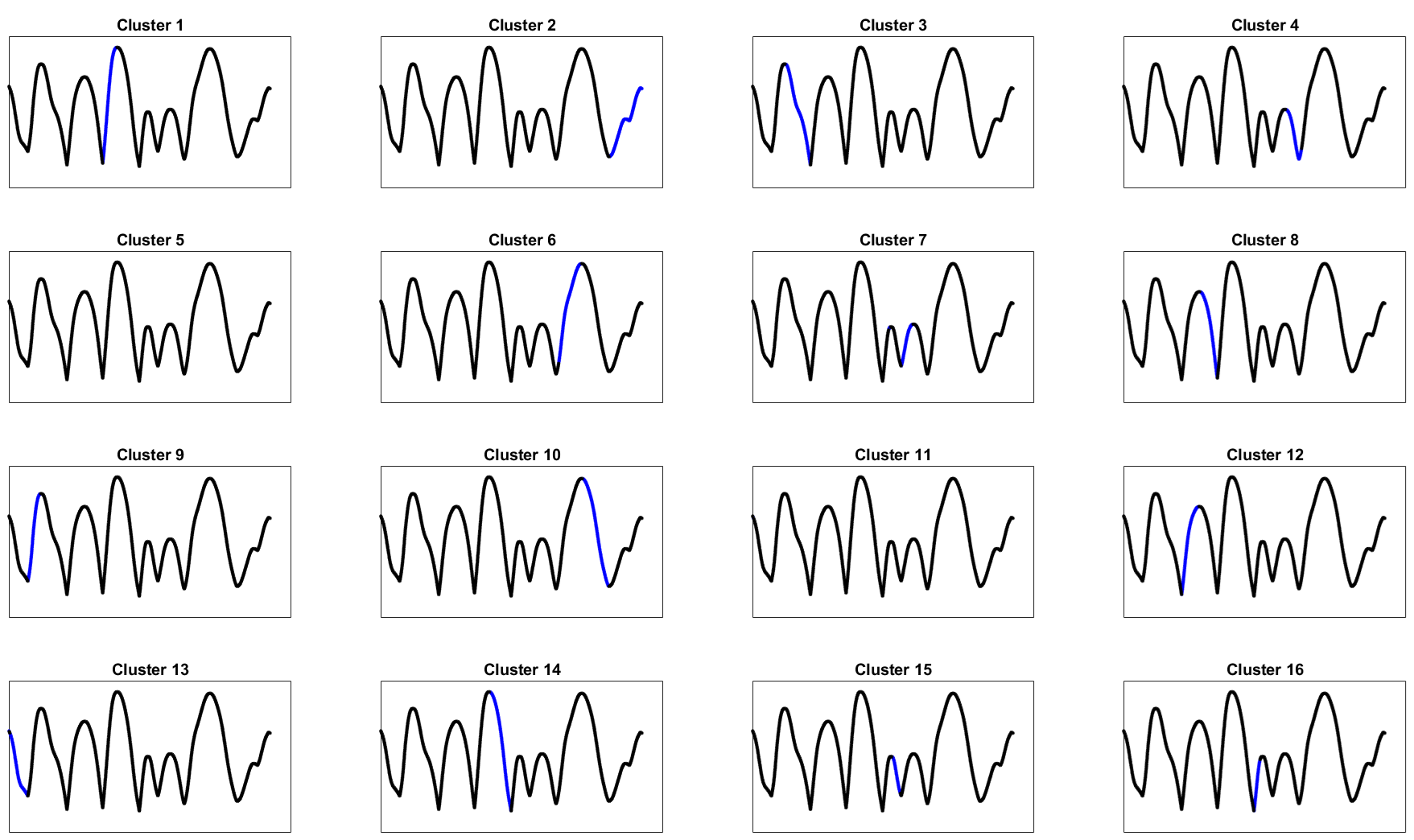}
\caption{Movement is decomposed in fragments of the $(t, v)$ components. Source: \cite{mazzetti2023functional}.}
\label{clusterini_mazzetti2}
\end{figure}

\begin{figure}[H]
\centering
\includegraphics[scale= 0.35]{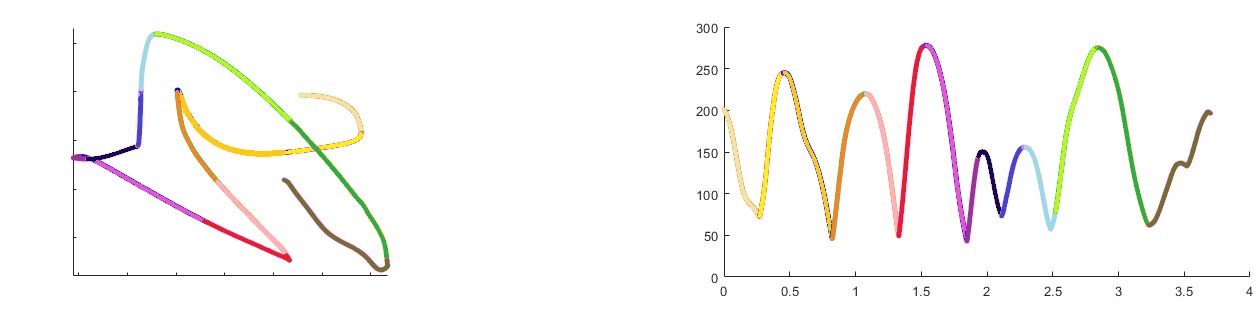}
\caption{Movement is decomposed in fragments, according to the decomposition from \cite{mazzetti2023functional}.}
\label{clusterini_mazzetti3}
\end{figure}

It is easy to verify that the vector fields $\left(X_i\right)_{i=1}^{3}$  together with their commutators satisfy the  H\"ormander condition, which can be stated as following: 
\begin{definition}\label{Hordef}
We say that a family of vector fields satisfy the H\"ormander condition if together with their commutators of any order, they span the whole tangent space at every point. 
\end{definition}
 When condition is fulfilled, it is always possible to prove that any couple of points can be connected with a horizontal curve. In \cite{mazzetti2023functional} it was also verified that couple of points can be connected by admissible horizontal curves. As a consequence, the distance is always bounded, and a good estimate of the heat kernel in the space is the following 
 function 
\begin{equation}\label{kernel_m}
K_{\mathcal{M}}\left(\eta_0,\eta\right)= e^{- d_{\mathcal{M}}\left(\eta_0,\eta\right)^2},
\end{equation}
where we used the letter $\eta$ to denote the general point $(x,y,t,\theta,v,a)$. 
This kernel represents the diffusion in the geometry of the space, so that it can describe the propagation of the signal in the cortical structure. For this reason it 
has been proposed as an estimate of the local connectivity weight between the cortical tuning points $\eta_0$ and $\eta$.  
In \cite{mazzetti2023functional} a spectral clustering algorithm based on  this kernel has been applied. The points are grouped in short curves, which have the properties of the fragments experimental found and in particular the fragments obtained in \cite{kadmon2019movement}(see Figures \ref{clusterini_mazzetti1}, \ref{clusterini_mazzetti2} and \ref{clusterini_mazzetti3}). 
However the algorithm does not group the fragments in neural states.

\section{A kinematic model of neural states}\label{sec3}
In the following we will obtain neural states via a grouping algorithm in the space of fragments.  

We observe that the classification of fragments in neural states obtained in \cite{kadmon2019movement}
 is invariant with respect to $(x,y)$. In order to model this property, we will consider a sub-manifold $\mathcal{M}_1$ independent of the variables $(x,y)$ of  the feature space $\mathcal{M}$ defined in \eqref{6D}, we will define a metric on $\mathcal{M}_1$, and use it as a pseudo metric on $\mathcal{M}$. The clustering with this pseudometric will be independent of $(x,y),$
and will provide a model for neural states
\subsection{A sub-manifold of the feature manifold}
 
Let us consider a  4D space 
subspace of $\mathcal{M}$, which does not depend on $(x,y)$

\[\mathcal{M}_1 = \{(x,y,t, \theta, v,a): x=y=0\}= \mathbb{R}^+_t\times  S_\theta^{1} \times \mathbb{R}^+_v \times \mathbb{R}_a,
\]

The vector fields $X_i$ defined in \eqref{campi_2D} can be restricted to the tangent plane to $\mathcal{M}_1$ (see also \cite{Ambrosio2006intrinsic, citti2006implicit}) and become 
\small
\[\hat X_{1}=  a\frac{\partial}{\partial{v}}+ \frac{\partial}{\partial{t}},\quad \hat X_{2}= \frac{\partial}{\partial{\theta}}, \quad \hat X_{3}= \frac{\partial}{\partial{a}}. \]

We will choose as horizontal distribution for  $\mathcal{M}_1$ the sub-bundle of the tangent bundle generated by these vector fields at every point, and we define on this distribution the metric which makes $(\hat X_{i})$ orthonormal. In this way $\mathcal{M}_1$ becomes a sub-Riemannian manifold.

Let us compute explicitly the commutators of these vector fields: 
\begin{equation}\label{Hei}
\hat X_4 = [\hat X_{1},  \hat X_{3}] = \frac{\partial}{\partial{v}},\end{equation}
while all the other commutators vanish.

Equation \eqref{Hei} shows that the H\"ormander condition (recalled in Definition \ref{Hordef}) is satisfied by the vector fields $(\hat X_i)_{i=1}^3$. 
We will define a dilation on the vector fields, calling 
\[\delta_\lambda(\hat X) = \lambda \hat X,\]
for every $\hat X\in span (\hat X_1, \cdots , \hat X_3)$. Consequently we will get 
\[\delta_\lambda(\hat X_4) = [\lambda \hat  X_1, \lambda \hat  X_2] = \lambda^2  [ \hat  X_1,  \hat  X_2]= \lambda^2    \hat  X_4 \]
Due to these dilation, it is natural to {\color{blue}assign} a different degree to elements of the tangent space. We will assign degree 1 to the vector fields $\hat X_i$, with $i=1,2,3$, while we will assign degree 2 to the vector field obtained as commutator:
\[deg(\hat X_i) =1 \text{ for } i=1, \cdots 3, \;\; deg(\hat X_4) =2.\]
Then we define a norm on tangent space, homogeneous with respect to these dilation: if $e = \sum_{i=1} ^4 e_i \hat X_i$, then we define  norm of $e:$
\begin{equation}\label{norm}
|e|  = \sum_{i=1} ^4 |e_i|^{1/ deg(X_i)}.
\end{equation}
This is homogeneous in the sense that $|\delta_\lambda (e)|= \lambda |e|$. Let us recall that the measure of the associated ball  
$B(0,r) = \{e: |e|\leq r\} $ is exactly

\begin{equation} \label{Hdim}|B(0,r) | = r^Q,{\text  where   }\; \;
Q= \sum_{i=1}^4 \text{deg}\left(\hat X_i\right) =5. 
\end{equation}
This is why we will call the constant $Q$ homogeneous dimension of the space (see for example \cite{nagel1985balls}). It is clear that $Q$ is greater than the topological dimension of the space, which is 4, and we will see that it plays a role in the estimation of the distance and the measure of the ball of the space. 
 Consequently we will define 
 
 We call a horizontal curve any integral curve in $\mathcal{M}_1$ of the vector fields ($\hat X_i)_{i=1}^3$, and the length of any horizontal curve $\gamma$ \begin{equation}
l\left(\gamma\right)= \int_0^1 \left|\gamma'\left(t\right)\right| dt,
\end{equation}
where $|\cdot |$ denotes the horizontal norm introduced on the distribution in \eqref{norm}. 

Since the H\"ormander condition is satisfied, the Chow Theorem\cite{chow1940systeme} ensures that any couple of points of the space can be joined by an integral curve 
of the vector fields $(\hat X_i)_{i=1}^3$.
Consequently,
we can define a distance between any couple of points 
$\hat \eta_0 =(t_0, \theta_0, v_0, a_0)$ and 
$\hat \eta =(t, \theta, v, a)$: 
\begin{equation}\label{cc}
d_{\mathcal{M}_1}\left(\hat \eta_0, \hat \eta\right)= \inf\left\{l\left(\gamma\right): \gamma\; \text{is an horizontal curve connecting}\: \hat \eta_0\; \text{and}\; \hat \eta\right\}.
\end{equation}

The curves on which the minimum is attained are called geodesics. 
Since 
the commutator relations \eqref{Hei} characterize the vector fields $X_1, \cdots  X_3$ as the generators of the Heisenberg Lie algebra, the explicit expression of geodesic distance is known, (see for example \cite{mazzetti2023model}). However, in order to simplify computation, we use here an estimate of the distance proved by \cite{nagel1985balls} in terms of exponential coordinates. Recall that the exponential map is defined as follows 
\begin{definition}
Let $X$ be a smooth vector field, and let $\hat \eta_0$ be a point in $\mathcal{M}_1$. We denote $exp(sX)(\hat \eta_0)$ the solution of the Cauchy problem 
\[\gamma' = X(\gamma(s)), \quad \gamma(0)= \hat \eta_0.\]
\end{definition}
The exponential mapping is a local diffeomorphism, and it induces a choice of coordinates
\begin{definition}\label{expcoord_def}
Let $\hat \eta_0\in\mathcal{M}_1$ fixed. We define canonical coordinates of $\hat \eta$ around a fixed point $\hat\eta_0$, the coefficients $e = (e_1, \cdots, e_4)$ such that 
\begin{equation}
\hat\eta= \exp\left(\sum_{i=1}^4 e_i \hat X_i\right)\left(\hat \eta_0\right).
\end{equation}
\end{definition}
A direct computation provides us the expression of the exponential map and the canonical coordinates $e_i$:

\begin{remark}\label{contiremark}
We will show that the expression of the canonical coordinates is the following 
\[e_{1}= t_{1}- t_{0}, \;\; e_{2}= \theta_{1}- \theta_{0}\;\;, e_{3}= a_{1}- a_{0}, \;\;   e_{4}= \left(v_{1}-v_{0}\right) -\frac{t_{1}- t_{0}}{2}\left(a_{0}+a_{1}\right).\]
\end{remark}
\begin{proof}
In order to obtain these expressions, we simply use the definition and we consider the system
\begin{align*}
\begin{cases}
\dot{\gamma}\left(s\right)=& e_{1}\hat X_{1} + e_{2} \hat X_{2}+ e_{3}\hat X_{3}+ e_{4}\hat X_{4}\\
\gamma\left(0\right)=& \left(t_0, \theta_{0}, v_{0}, a_{0}\right)\\
\gamma\left(1\right)=& \left(t_1, \theta_{1}, v_{1}, a_{1} \right),\\
\end{cases}
\end{align*}
and we get 
\begin{align*}
\dot{\theta}=  e_{2}, \;\; 
\dot{v}=  e_{1}a + e_{4}, \;\; 
\dot{a}=  e_{3}, \;\; 
\dot{t}=  e_{1}.
\end{align*}

In this way we also get $v\left(s\right)= e_{1}e_{3}\frac{s^{2}}{2} + e_{1}a_{0}s + e_{4}s+ v_{0}$ and consequently $ e_{4}= \left(v_{1}-v_{0}\right) -\frac{e_{1}}{2}\left(a_{0}+a_{1}\right)$.
\end{proof}

A local estimate of the distance have been obtained in large generality in \cite{nagel1985balls}, hence the following estimate holds for $\hat{\eta}_0$,$\hat{\eta}_1$ in a compact set $K$.

\begin{proposition}
For every compact set $K$ there exist constants $C_0, C_1$ such that for every $\hat{\eta}_0$,$\hat{\eta}_1 \in K $ the distance defined in \eqref{cc} between them locally satisfies 
\begin{equation}\label{distanza_exp_6D}
C_0 \, d_{\mathcal{M}_1}\left(\hat{\eta}_0,\hat{\eta}_1\right)
\;\leq\; 
|e|\;\leq\;
C_1 \, d_{\mathcal{M}_1}\left(\hat{\eta}_0,\hat{\eta}_1\right),
\end{equation}
where $e$ are the canonical coordinates of 
$\hat{\eta}_1$ around $\hat{\eta}_0$, as defined in Definition \ref{expcoord_def}. 
\end{proposition}


 \bigskip
 
 The commutation relations \eqref{Hei}, characterize the Heisenberg Lie algebra. Precisely the variables $t,v, a,$ the space can be identified with the elements of the Heisenberg group $H^1$, while the variable $\theta$ belongs to the group $S_\theta^{1}$. Consequently the whole manifold $M_1$ coincides with 
$H^1\times S_\theta^{1}$. This allows to estimate separately the distance restricted to $H^1$ and the component $\theta$ in $S_\theta^{1}$. 
The exponential map in the Heisenberg group is a global diffeomorphism, so that the distance can be defined globally with the same expression. This is not the case for $S_\theta^{1}$: in this set formula \eqref{distanza_exp_6D}, only provides a local estimate, but of course the angle is periodic, hence we will replace $e_2$ by 

\[\hat e_2 = 2\sin\Big((\theta_0 - \theta_1)/2\Big),\]

which has the same behavior in $0$, but the required global periodicity. The distance can now be estimated by 
\begin{equation}\label{distanza_periodic}
\left(\left|e_{1}\right|^{2} + \left|\hat e_{2}\right|^{2}+ \left|e_{3}\right|^{2}+ \left|e_{4}\right| \right)^{\frac{1}{2}}.
\end{equation}

\subsection{A pseudo-metric in the space of features}

As recalled at the beginning of Section \ref{sec3}, we are interested in a classification of curves which is independent of variables $(x,y).$
Hence we need to introduce on 
$\mathcal{M}$
a pseudo-distance independent of these variables, and this will be made by means of the distance defined on ${\mathcal{M}_1}$.

Precisely  the distance $d_{\mathcal{M}_1}$ can be extended on $\mathcal{M}$
simply setting 
\[d_{\mathcal{M}_1}\Big((x,y, t, \theta, v, a ),  (x_0,y _0, t_0, \theta_0, v_0, a_0 )\Big) =d_{\mathcal{M}_1}\Big((0,0, t, \theta, v, a ),  (0,0, t_0, \theta_0, v_0, a_0 )\Big) .\]

Clearly, this function will vanish on a couple of points with the same components $t, \theta, v, a $ and different $(x,y)$ components.  
Consequently, it is not a distance, but a pseudo-distance. 
Indeed the notion of pseudo-distance is the following

\begin{definition}
A pseudo-metric space 
$(\mathcal{M}_1 ,d)$ is a set 
$\mathcal{M}_1$ together with a non-negative real-valued function 
$d: \mathcal{M}_1\times \mathcal{M}_1 \to \mathbb {R}^+ $ called a pseudo-metric, which satisfies \[d(\eta, \eta) = 0\text{  for every }\eta \in \mathcal{M}_1,\]
is symmetric and satisfies the triangle inequality. 
In particular $d(\eta, \eta_0) = 0$ does not imply in general that $\eta=\eta_0$.
\end{definition}

\subsection{A pseudo-metric in the space of fragments and cortical connectivity}

We model 
fragments as horizontal curves defined on the same time interval $[0,1]$ 
with values in the 6D space $\mathcal{M}$ introduced in \eqref{M}. Precisely, if $X_1, X_2, X_3$ are defined in \eqref{campi_2D} then the space horizontal curves is defined as 
\[\mathcal{H}=\{\gamma: [0,1] \to \mathcal{M}: \gamma' (s) = \alpha_1X_1 + \alpha_2 X_2+ \alpha_3 X_3: \alpha_i \;\; \text{
are regular functions }\}.\]
We have already introduced the space of fragments as a subset of the set of horizontal curves. The coefficients $\alpha_1$ will be 1, the coefficient  $\alpha_2 $ of the vector field $X_2$ can be choosen to be constant, while the
choice of $\alpha_3$  ensures that the acceleration vanishes at the initial and final
point and has a bell-shaped graph 
(see also \eqref{eqa}, for more detailed explanation). 

\begin{align}\label{fragments}
\mathcal{F}=\Big\{\gamma: [0,1] \to \mathcal{M}: \gamma' (t) = X_1 + \alpha_2 X_2+ \alpha_3 X_3:   \quad \quad \quad\quad \quad \quad \\
\quad \quad \quad\quad \quad \quad \gamma(0) = \eta_0 \in \mathcal{M}, \alpha_1, \alpha_2, j \in \mathbb{R}, \;\; \alpha_3(t) =  j\Big(t- \frac{1}{2}\Big)\;\Big\}.\nonumber
\end{align}
Note that this space is finite-dimensional, since it depends only on the parameters $\eta_0 \in \mathcal{M}, \alpha_1, \alpha_2, j \in \mathbb{R}$. 
In this space we want to apply a new clustering algorithm to find neural states, hence we introduce a suitable pseudo-distance on it. 
   The pseudo metric $d_{\mathcal{M}_1}$ defined on the space $\mathcal{M}$ naturally defines a pseudo-distance in the space $\mathcal{F}$.

 \begin{definition}
If $\gamma_1, \gamma_2 \in \mathcal{F}$, then we can call 
\[ d_\mathcal{F}(\gamma_1, \gamma_2) = \int_0^1 ||\gamma'_1(t)-\gamma_2'(t)||_{\mathcal{M}_1} dt +
d_{\mathcal{M}_1}(\gamma_1(1),\gamma_2(1)).
\]
 \end{definition}

 \bigskip 
 Let us explicitly note that this is a pseudo-distance, as the pseudo-distance between two curves obtained via translation is 0. 



\bigskip

In analogy of what was proposed in section \ref{sec2}, we introduce here a kernel, starting with a local approximation of the heat kernel at fixed time. The heat kernel in the space of fragments will model the propagation of the signal along connectivity in the space of fragments. According to \cite{nagel1985balls} it can be locally estimate  as follows:

    \[
    K_\mathcal{F}: \mathcal{F} \times \mathcal{F} \to \mathbb{R}\]
    \begin{equation}\label{connectivityF}
        K_\mathcal{F}(\gamma, \bar \gamma) = e^{-d_\mathcal{F}(\gamma, \bar\gamma)^2},
    \end{equation}

for every couple of curves $\gamma, \bar \gamma$. 
\subsection{Cortical activity}

The evolution of the neuronal population activity has been classically modeled through a mean field equation firstly proposed in the works of Amari \cite{amari1972characteristics} and Wilson and Cowan \cite{wilson1972excitatory}, and largely developed in literature (see \cite{ermentrout1980large, bressloff2003functional, faye2010theoretical, sarti2015constitution}). In the space of fragments, it is expressed in terms of the connectivity kernel as follows: 
\begin{equation}
\label{eq:meanField}
\frac{du(\gamma, s)}{ds}= - \mu_1 u (\gamma, s)+ \mu_2\varrho\left(\int_{\Omega} K_{\mathcal{F}}(\gamma, \gamma')u(\gamma', s)d\gamma' + h(\gamma, s)\right),
\end{equation}
where $s > 0$, the coefficients  $\mu_1$ and $\mu_2$ represent the decay of activity, and a short-term synaptic facilitation respectively, 
The function $\varrho$ is the activation function, typically a sigmoid or a ReLU, and $h$ is the input. We explicitly note that the integral is extended on a space of curves, but the space of fragments is parametrized via a finite number of parameters, which reduces the integral to a standard finite-dimensional one.  

In the definition of the domain we will follow an approach introduced in \cite{sarti2015constitution}. The integration can be restricted to the set where the activity $a$ does not vanish, which is the set of points activated by the stimulus. If the feedforward input $h$ can attain only two values, namely $0$ and a constant value $c$, and the strenth of connectivity is weak,  no new points are activated, so that  the domain reduces to
\begin{equation}
\Omega = \{ \gamma: h(\gamma) = c \};
\end{equation}

The stability of neural states can be studied  by mean of the eigenvalue problem obtained by linearizing the operator, and considering its time independent counterpart: 
\begin{equation}
\label{meanFieldEigv}
L u := -\mu_1 u + \varrho'(0) \mu_2 \int_{\Omega} K_{\mathcal{F}}(\gamma')u(\gamma', t)d\gamma' = \lambda u \;\;\iff\;\;\int K_{\mathcal{F}}(\gamma')u(\gamma')d\gamma'= \tilde \lambda u, \;\;
\end{equation}
with $\tilde \lambda =\frac{\lambda + \mu_1}{\gamma \mu_2} $.
For this reason, stable neural states can be studied 
in terms of a spectral analysis of the connectivity kernel. 
This argument has been developed in the paper \cite{sarti2015constitution} with the scope of finding a strict link between emergence of patterns in the brain,  and spectral clustering algorithms.

\subsection{Neural states obtained via grouping in the space of fragments}
We will use a spectral analysis technique of the connectivity kernel $K_{\mathcal{F}}$ defined in \eqref{connectivityF} to obtain emergence of neural states. To this end we consider a matrix $A$, discretization of the connectivity kernel 
\begin{equation}\label{affini}
A = a_{ij}= e^{-d_{\mathcal{F}}^2\left(\eta_i,\eta_j\right)},
\end{equation}
where $d_{\mathcal{F}}$ is a distance over the considered space.
It has been originally shown by Perona \cite{perona1998factorization} that the first eigenvector of $A$ can represent the first emergent pattern.
To reduce error due to noise, the affinity matrix can be suitably normalized.
Many normalizations have been proposed (e.g. \cite{butler2006spectral, ng2001spectral, shi2000normalized}): one of the most widely applied is the one presented by Meila and Shi \cite{meilua2001random} since it reveals properties of the underlying affinity matrix by ways of the
Markov chain, providing a probabilistic foundation of the clustering algorithm. Indeed,
the authors defined a Markov-type matrix P as follows
\begin{equation}
P = D^{-1}A, \quad D\; \text{diagonal matrix, }\quad d_i = \sum_{j=1}^n a_{ij}.
\end{equation}
The eigenvalues of $P$ are real, positive and smaller than one, while the eigenvectors have real components, and they are a basis of the space. Consequently we can represent fragments either in the original $\mathbb{F}$  space , or through the projection on eigenvectors, and there is a natural change of variables between these two representation. It has been proved in (see diffusion map theorem  \cite{lafon2006diffusion, coifman2006diffusion}) that Euclidean distance in the coordinates induced by the projection onto the eigenvectors is
equivalent to the distance $d_{\mathcal{F}}$ in fragment space. This means that
the clustering result obtained by applying the k-means algorithm to the data projected
onto the eigen space will be the same as the result obtained by applying the k-means
algorithm to the data directly in the affinity matrix space. For this reason, a $k$-means algorithm in this coordinates will provide the classification for our problem. In particular, we apply here to our kernel and its discretizations provided in \eqref{affini}, a simple and efficient algorithm has been proposed in \cite{kannan2004clusterings}:
\begin{enumerate}
\item Starting with the previous defined affinity matrix, calculate the normalized affinity matrix $P = D^{-1}A$ (skip this step if $A$ is a block diagonal matrix).
\item Solve the eigenvalue problem $PU= \lambda U$, where $U$ is the matrix formed by the column eigenvectors $\lbrace u_i\rbrace_{i=1}^n$.
\item Find the eigenvectors whose eigenvalues are over a fixed threshold, i.e. find $\lbrace u_i\rbrace_{i=1}^q$ such that $\lbrace \lambda_i\rbrace_{i=1}^q> 1- \epsilon$.
\item Assign the data set points to the cluster with an Euclidean clustering algorithm.
\end{enumerate}

\section{Results}\label{sec4}

\subsection{Test on uniformly generated data}
\begin{figure}[H]\label{f1}
\centering
\includegraphics[width=9 cm]{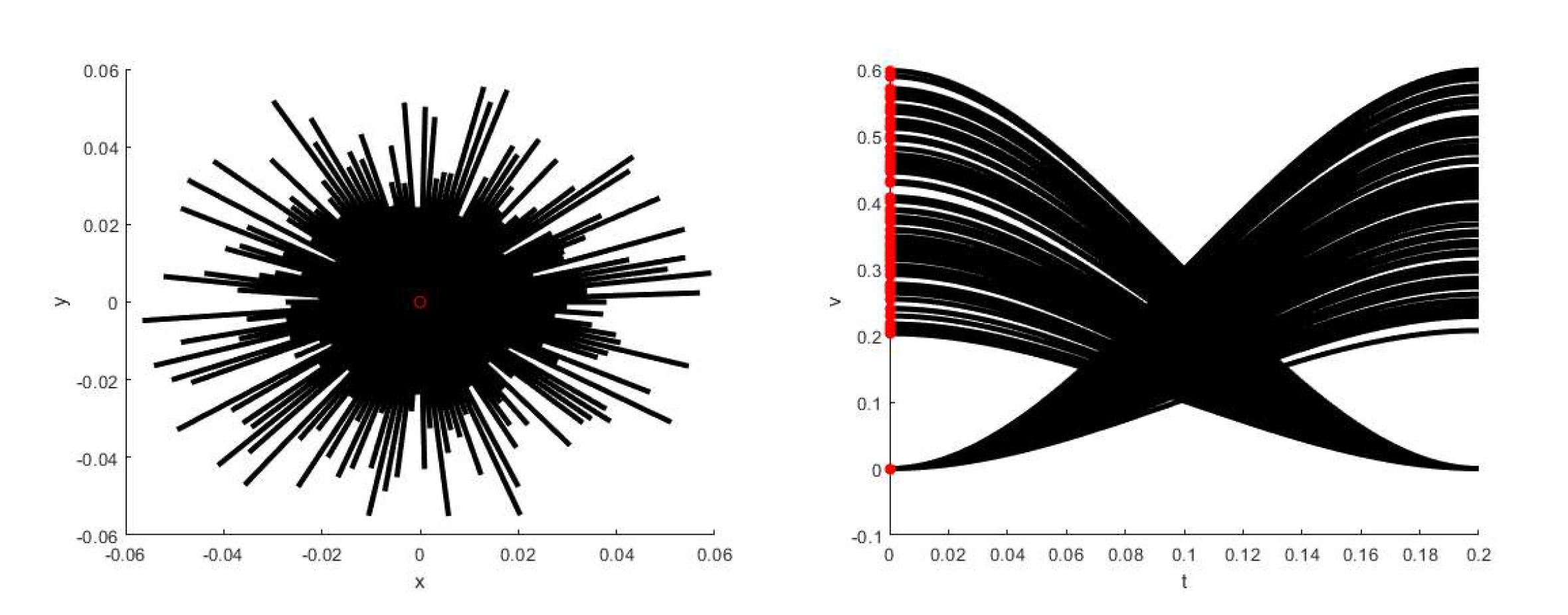}
\caption{A family of curves starting from the origin with constant direction $\theta$ (left image) and derivative of the acceleration $j$ uniformly distributed (right).}
\label{constant}
\end{figure}

 We start by testing our model on samples of curves generated by the expression 
of fragments(see Figure \ref{constant}) introduced in \eqref{fragments}. 
Each fragment depends on 9 variables: the initial position $\eta_0 = (x_0, y_0, t_0, \theta_0, v_0, a_0)$, and the coefficients $\alpha_1, \alpha_2, j$
in \eqref{fragments}. 
To simplify visualization we choose in a first example the initial position of all trajectories at the origin: $x_0=0, y_0=0, t_0=0$, 
$\theta_0$ uniformly distributed in $[0,2\pi]$, $v \geq 0$, $\alpha_1 =1$, $\alpha_2=0$ and $j$ uniformly distributed.

\begin{figure}[H]
\centering
\includegraphics[width=11 cm]{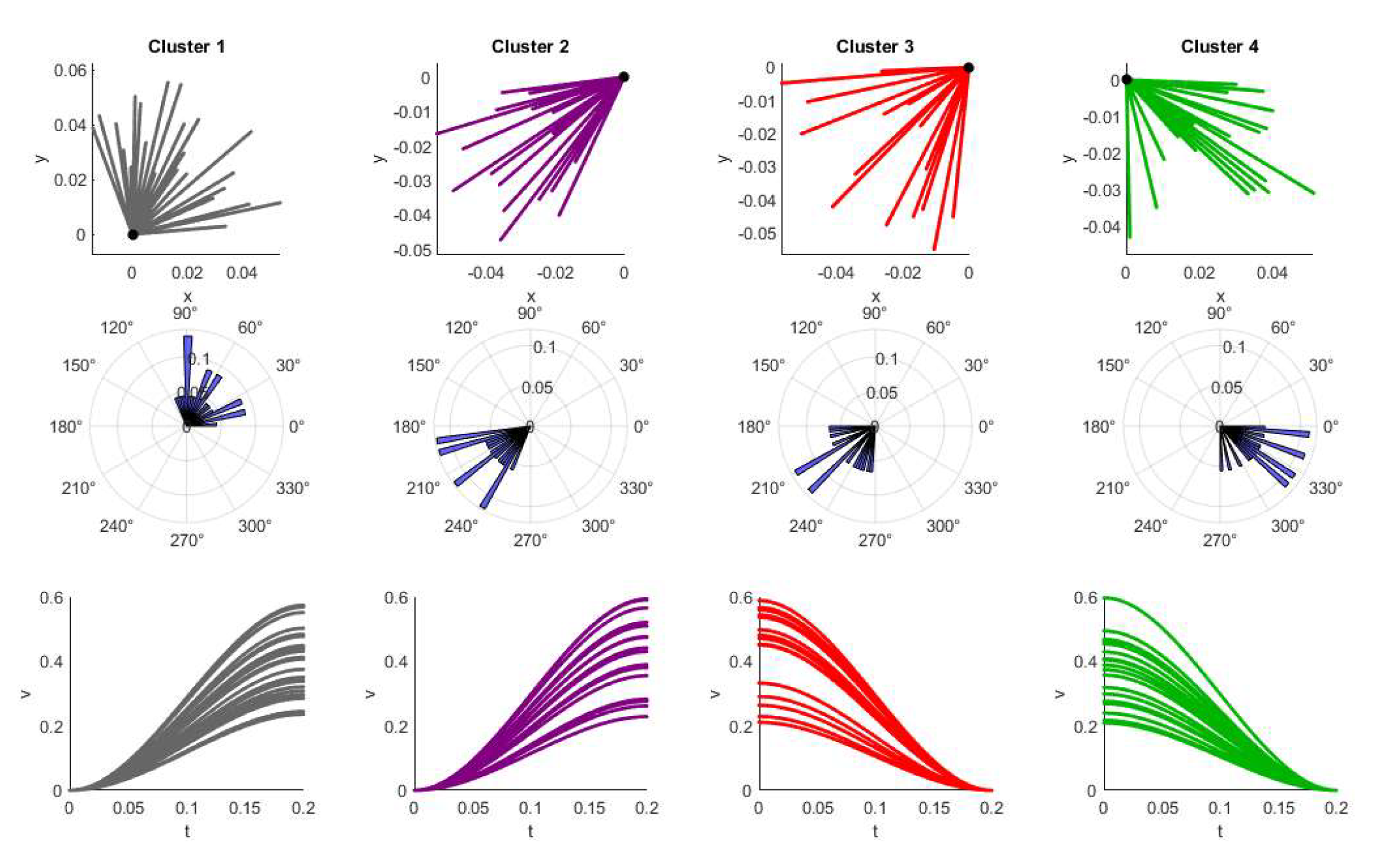}
\caption{A visualization of the grouping of fragments in neural states: the first 4 states. For each state, we visualize the $(x,y)$ projection (first row), the mean orientation (second row) and the projection in the $(t,v)$ plane(third row)}\label{f2}
\end{figure}

\begin{figure}[H]
\centering
\includegraphics[width=11 cm]{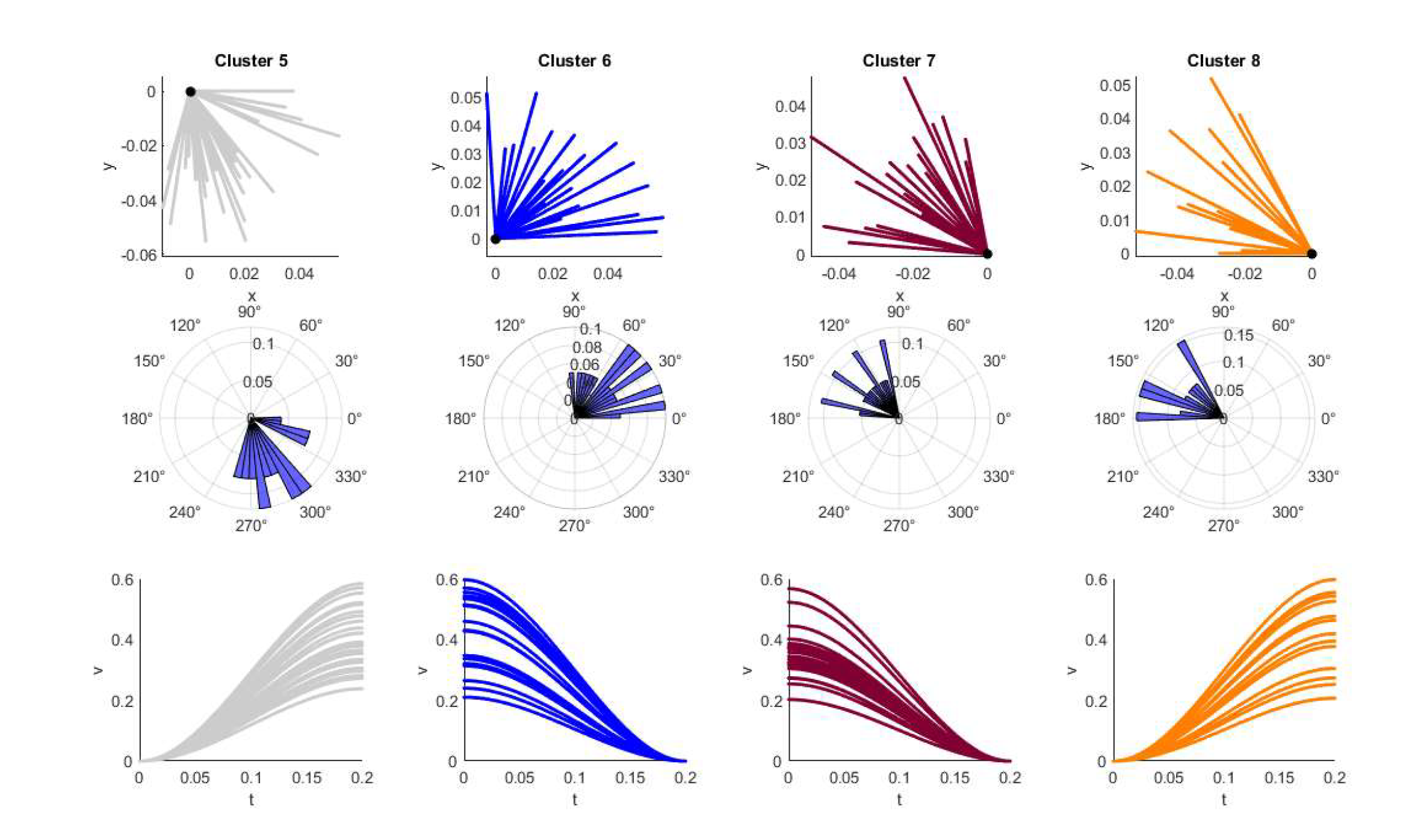}
\caption{A visualization of the grouping of fragments in neural states: the last 4 states. As before for each state we visualize the projection on the $(x, y)$  plane (first row), the mean
orientation (second row) and the projection in the $(t, v)$ plane (third row)}\label{f3}
\end{figure}

We apply the clustering algorithm and in this case, we obtain a correct clustering of the curves, in eight clusters, each one characterized by the orientation belonging to a specific quadrant and increasing or decreasing velocity  (see Figures \ref{f2} and \ref{f3}). 

\subsection{Test on randomly generated data}
 
\begin{figure}[H]\label{f4}
\centering
\includegraphics[width=9 cm]{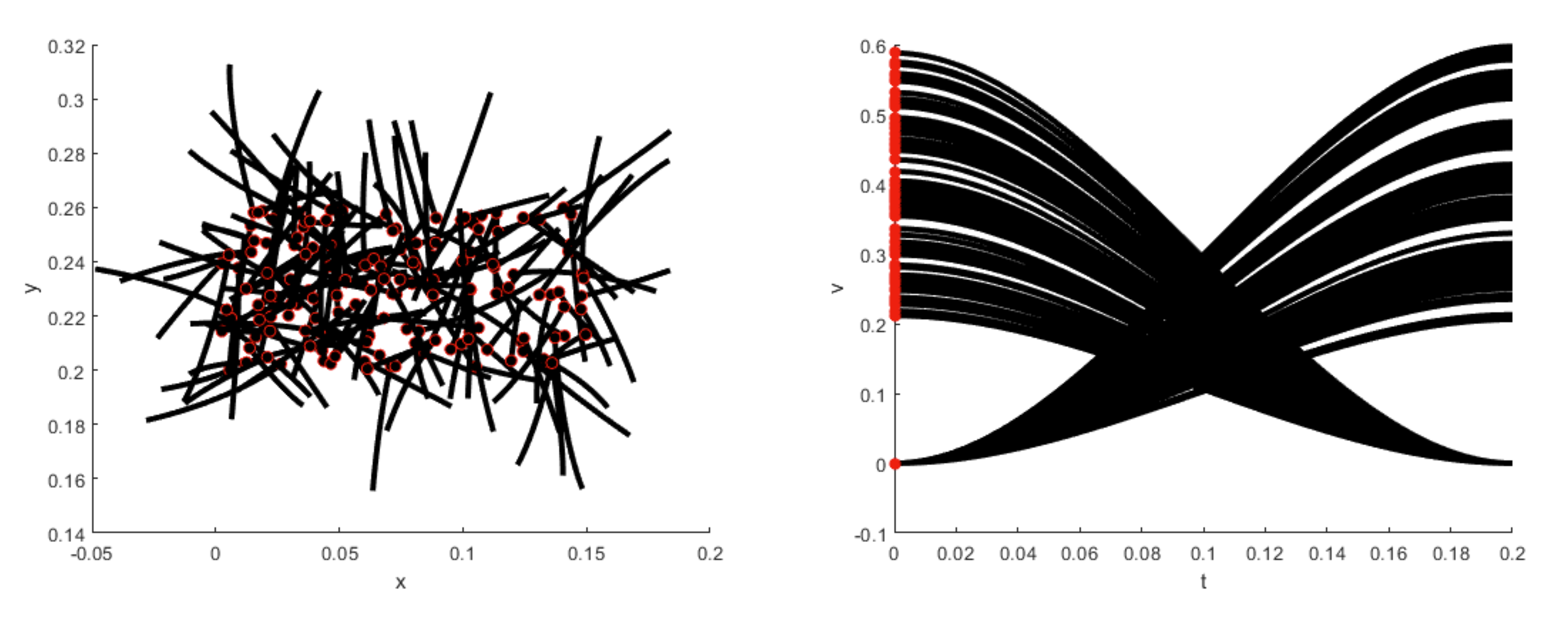}
\caption{A family of curves with all parameters randomly choosen}
\label{random}
\end{figure}

\begin{figure}[H]\label{f5}
\centering
\includegraphics[width=11 cm]{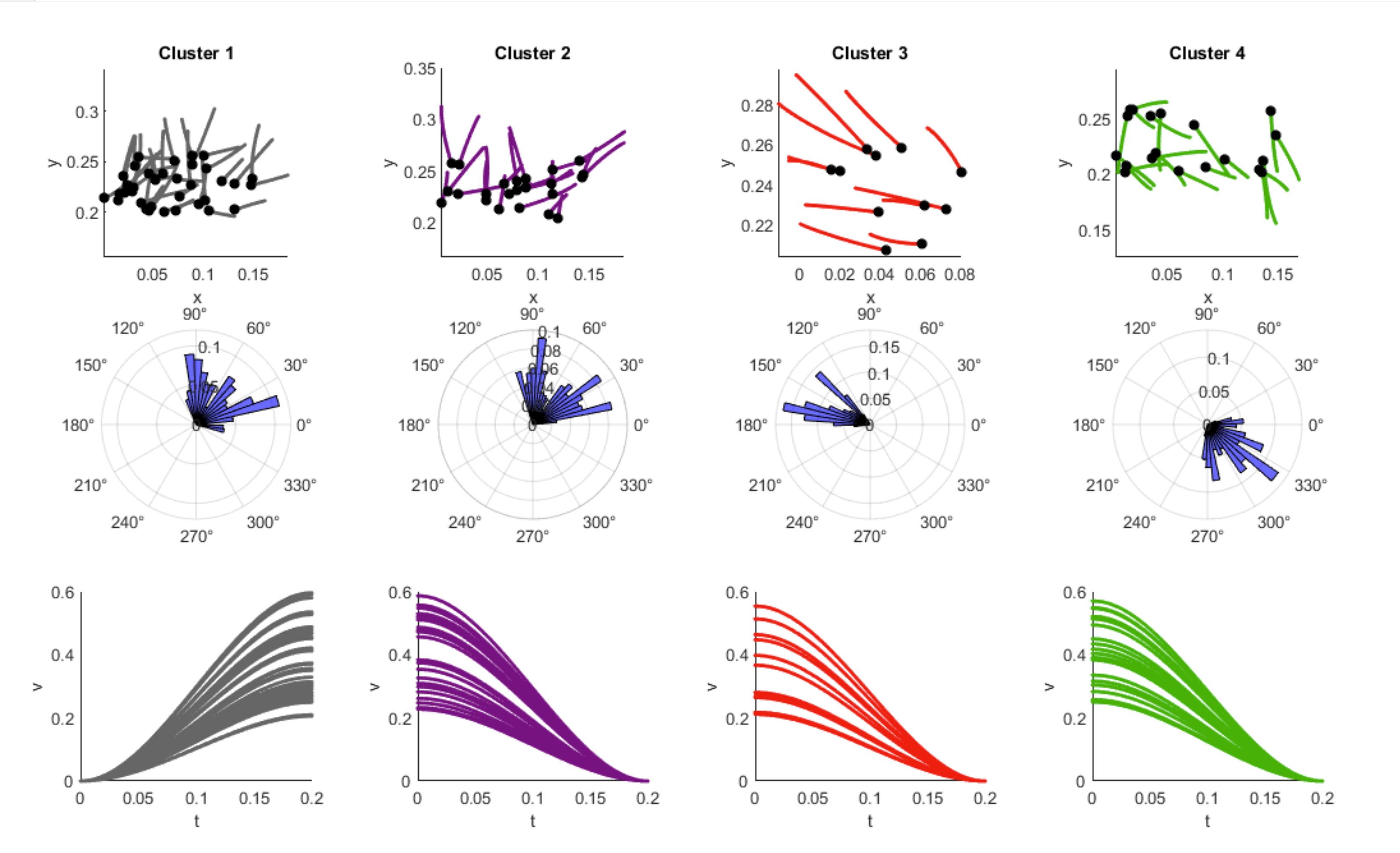}
\caption{Visualization of the grouping of fragments in neural states: the first 4 states. The same convention as in Figure \ref{f2} is adopted for visualization.}
\label{001}
\end{figure}

\begin{figure}[H]\label{f6}
\centering
\includegraphics[width=11 cm]{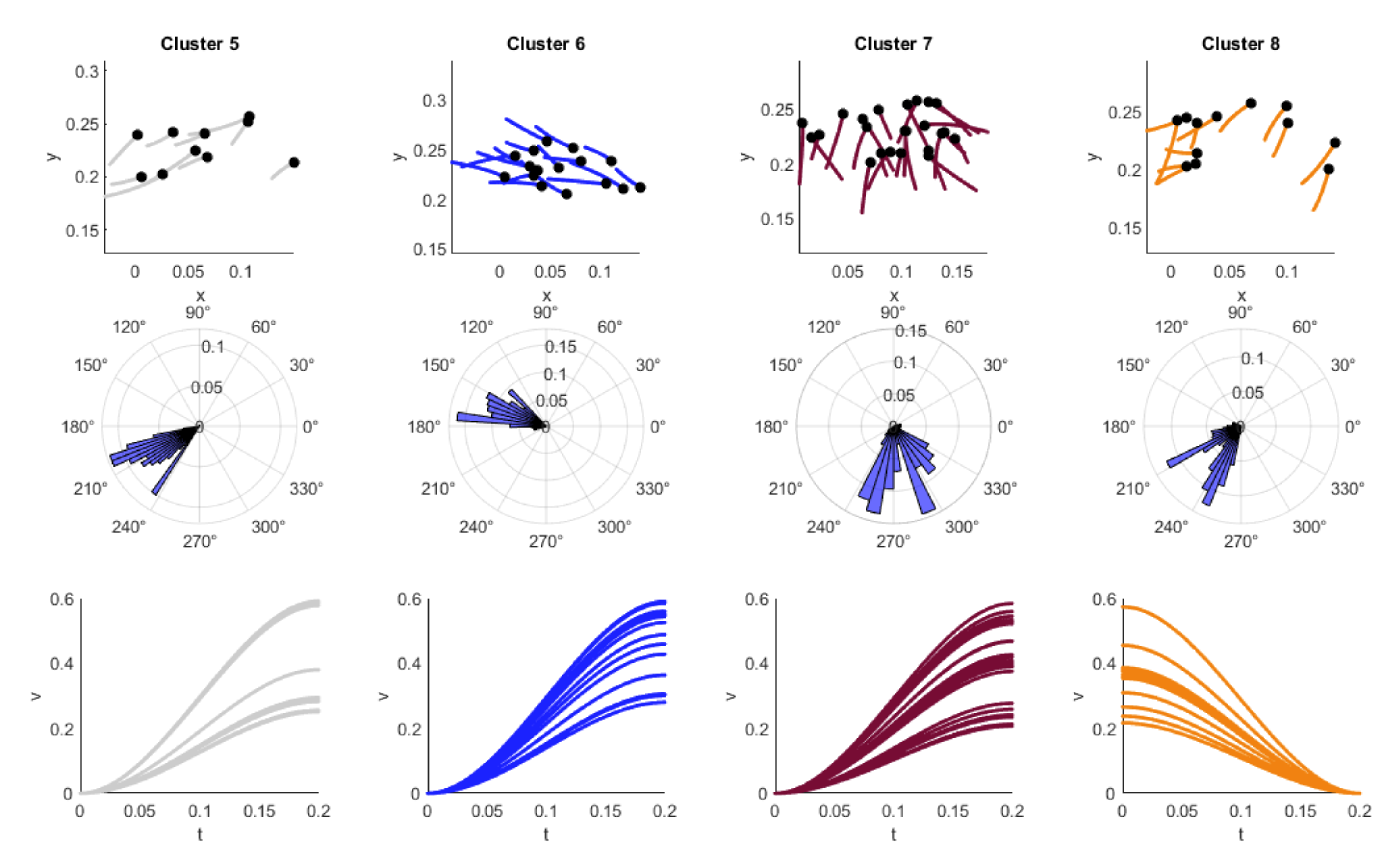}
\caption{Visualization of the grouping of fragments in neural states: the last 4 states. The same convention as in Figure \ref{f2} is adopted for visualization.}
\label{002}
\end{figure}

We test now the model on fragments 
defined as in \eqref{fragments}, with all parameters randomly chosen (see Figure \ref{random}). 
Also in this case we obtain a correct clustering of the curves(see Figures \ref{001} and \ref{002}).

\bigskip

\section{Conclusion}\label{sec5}

We introduced a geometric model of the arm area of the motor cortex: this area codes complex motor primitives: from simple features, as direction of movement, to short trajectories of the hand, called fragments, to more complex patterns, which we will call here neural states. 

Here we model the space of fragments as a space of short curves with values in a space of kinematic parameters, introduced in \cite{mazzetti2023functional}, and we introduce a geometric kernel as a model of cortical connectivity, and we use it in a differential equation to express cortical activity. Applying a grouping algorithm to this model of cortical activity we recover the same neural states obtained in \cite{kadmon2019movement}, who applied a grouping algorithm on measured cortical activity. This suggests that the choice of these variables we made here is sufficient to explain this phenomenon and the distance we consider is the correct one to model cortical connectivity. 

The interest of the model relies in its modularity, which mimics the structure of the brain. Indeed a first grouping algorithm is applied in the space $\mathcal{M}$, and the emerging groups are identified as points in a more abstract space. This approach would like to mimic the behavior of the cells in the brain, which process the stimulus at higher and higher scales, to extract both local and global properties.


\end{document}